# Visualizing beat phenomenon between two amplitude-modulated (AM) light beams on a solar cell using smartphones.

*Santiago Ortuño-Molina, Francisco M. Muñoz-Pérez, Juan C. Castro-Palacio, Juan A. Monsoriu*

Centro de Tecnologías Físicas, Universitat Politècnica de València, Camino de Vera, s/n, 46022 València, Spain

The beat phenomenon arises when two harmonic oscillations of slightly different frequencies ($f_1$ and $f_2$) are superimposed. This phenomenon has been extensively exemplified in Physics teaching journals for different contexts. For instance, the acoustic beats between sound waves have been studied with tunning forks[1,2], conventional speakers[3] and microphones[3] and, more recently using smartphones[4,5]. Likewise, the mechanical beat has been studied by using a spring-mass system[6,7] or a pendulum[8]. The beat can also be visualized on the oscilloscope screen by superimposing electrical signals[9]. In relation to light waves, there are also reported works in the literature[10,11]. More exotic cases are, for instance, the "beating" phenomenon in the oscillation of the synodic period of the Moon[12] or the beats displayed with foam packing[13].

In this work, we present an interesting new case of beat phenomenon which involves light waves and electrical signals. The beat is produced when two amplitude-modulated (AM) light beams are incident on a solar cell. The signal measured with the solar cell when both lamps are switched on can be expressed as

$$A_1\sin(2\pi f_1 t + \varphi_1) + A_2\sin(2\pi f_2 t + \varphi_2) = A_E(t)\sin\left(2\pi \frac{f_1+f_2}{2} + \varphi\right), \tag{1}$$

where

$$A_E(t) = \sqrt{A_1^2 + A_2^2 + 2A_1 A_2 \cos(2\pi|f_1 - f_2|t + \varphi_E)}. \tag{2}$$

It can be seen that the frequency of the envelope, $A_E(t)$, which is the frequency of the beat, is $f_{\text{beat}} = |f_1 - f_2|$.

## The experiment

Figure 1 shows a photo of the experimental setup used in this work. It includes two smartphones used as signal generators (labels 1 and 2). These mobile phones are connected in series to two LED lamps (labels 4 and 5) and to a set of 4 batteries (6 V in total) (labels 7 and 8). Each lamp includes a panel with 24 LEDs. The amplitude-modulated beams from the lamps are incident on a solar cell (label 6) which is connected to a third mobile phone used as an oscilloscope (label 3). The signals are taken out and in the smartphones through the audio port by using audio jacks (labels 9, 10, 11). The solar cell (60 mm x 80 mm) is made of polycrystalline silicon. Its maximum output voltage is 1.5 V and 0.65 W of power.

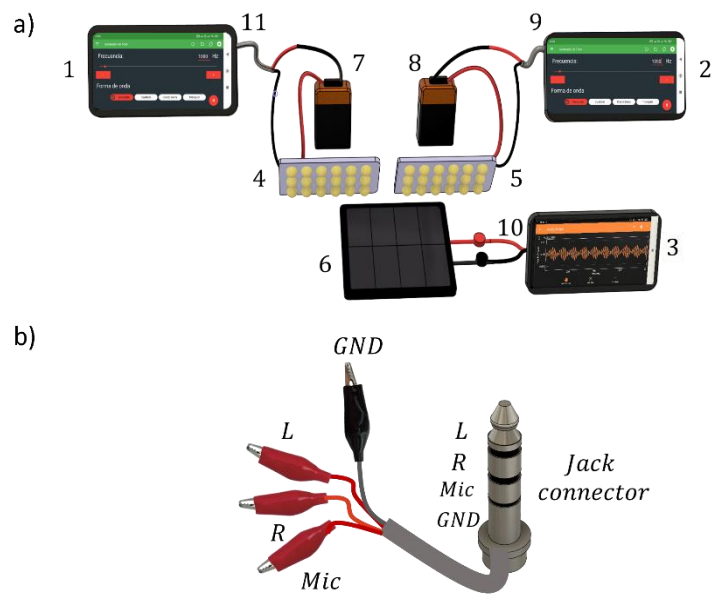

**Fig. 1.** Schematic representation of the experimental setup used in this work to visualize the beat phenomenon between two light beams (panel a) and of the jack connector used (panel b).

Smartphones have been used in previous work as signal generators and as oscilloscopes where the signals are sent in and out through the audio port by means of a jack connector. Recent works on RC[14], RL[14], and RLC[15] series circuits have proven these possibilities of smartphones in Physics teaching laboratory. For instance, the smartphone can be turned into a signal generator by using the Tone Generator option available in the free app for Android Physics Sensors Toolbox Suite[16]. Similarly, it can be turned into an oscilloscope by using the Audio Scope option included in the free Android app phyphox[18].

Using the experimental setup in Figure 1, two sine waves with frequencies 1000 Hz and 1050 Hz, have been generated with the smartphones labelled 1 and 2. For this purpose, the Tone Generator option of the free app Physics Sensors Toolbox Suite[16] haven been used. The resulting variable voltage with the same frequency at the audio port of each smartphone is taken out with audio jacks and used to power the lamps along with a set of four batteries connected in series. The amplitude-modulated (or intensity-modulated) light beams are incident on the solar cell which is connected to a third smartphone (label 3) by means of the audio port and an audio jack. The photocurrent generated by a solar cell is proportional to the incident light intensity. Even when the measured voltage with the smartphone is expected to have a logarithmic dependence with the photocurrent, the experimental results show that the relationship between these quantities can be assumed nearly linear, which has to do with the fact that the actual voltage and current variations are small. The resulting beat is then visualized on the screen of the third smartphone using the Audio Scope option of the free Android app phyphox[17]. A Samsung Galaxy A52s smartphone and a Xiaomi Redmi Note 11 smartphone were used as signal generators and a Redmi Note 7 smartphone as an oscilloscope.

## Results

Figures 2 and 3 show the exported data as registered with the third smartphone used as an oscilloscope. The light intensity collected with the solar cell is plotted *versus* time in both graphs. Figure 2 shows the data of light intensity as collected with the solar cell for each of the two lamps independently. Fig. 2 a) corresponds to the lamp powered with a 1050 Hz signal and Fig. 2 b) to the one powered with a 1000 Hz signal. As commented above, the variable voltage signals are generated with smartphones. A quick calculation can be performed from the data shown on the plots. For the upper panel, we can see 10.5 oscillations in 0.01 s which corresponds to 1050 Hz of frequency. Similarly, for the lower panel, 10 oscillations in 0.01 s can be observed, which yields a frequency of 1000 Hz.

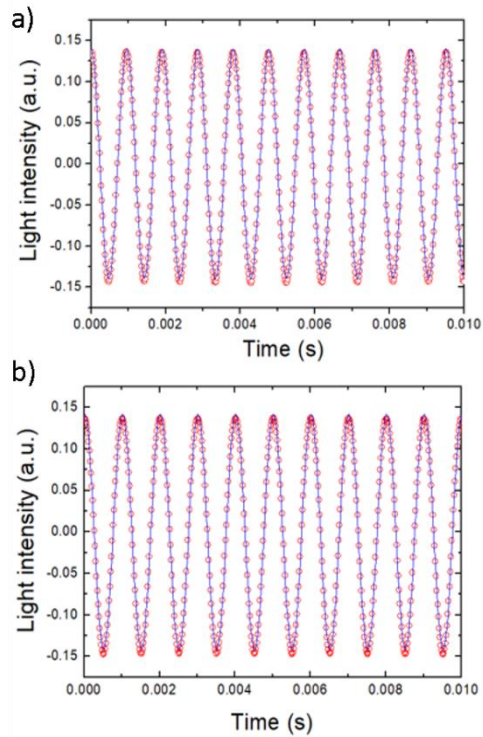

**Fig. 2.** Light intensity collected with the solar cell *versus* time for each lamp independently. a) shows the data for a signal of 1050 Hz and the b) for a signal of 1000 Hz.

These values can be derived more accurately from non-linear fittings (blue lines in the plots) using a sine function of the form $I(t) = I_m \sin(2\pi f t + \varphi)$, where $I(t)$ is the light intensity at each time as collected with the solar cell, $I_m$ the amplitude of the intensity and $f$ the frequency. For the data of the first lamp in the upper panel the resulting frequency is $(1050.031 \pm 0.057)$ Hz (and $R^2 = 0.9990$) and for the data in the lower panel $(999.952 \pm 0.068)$ Hz (and $R^2 = 0.9986$). The percentage difference with respect to the frequences set at the smartphones as signal generators is lower than 0.005 %.

On the other hand, figure 3 shows the exported data of the beat (when both lamps are switched on). A larger timescale was chosen for this plot in order to appreciate the beat oscillations properly. Table 1 (column 1) includes the times for the amplitude maxima of the envelope in figure 2. The time between two maxima (of the envelope) is shown in column 2. These times are used to calculate the period of the beat as $T = \langle \Delta t \rangle = (20.00 \pm 0.29)$ ms, where $\langle \Delta t \rangle$ is the average time over the times between consecutive maxima of the envelope. For the uncertainty of the period, the time between two consecutive maxima in the inner oscillations to the envelope is taken. The reported frequency of the beat is then $f_{\text{beat}} = \frac{1}{T} = (50.01 \pm 0.72)$ Hz. The uncertainty of the frequency of the beat is obtained by doing uncertainty propagation. It can be seen that there is an excellent agreement between this frequency and the $f_{\text{beat}} = |f_1 - f_2| = 50$ Hz as predicted by theory in Eq. (1).

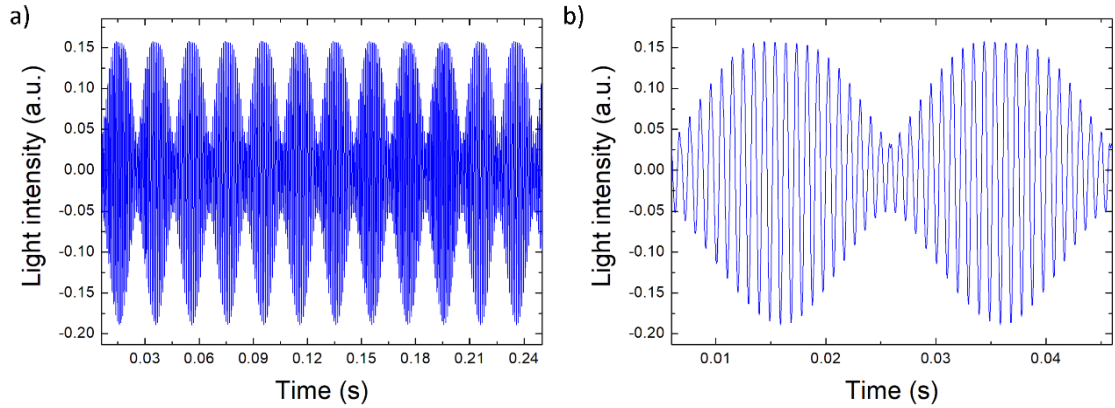

**Fig. 3.** Light intensity collected with the solar cell *versus* time for the beat between the intensity-modulated light beams of both lamps (panel a). To better observe the beat phenomenon, the first two oscillations are shown in panel b.

**Table 1.** Times for the amplitude maxima in figure 2 as registered with the smartphone. Column 3 indicates the time between two maxima which is a good measure of the period of the beat.

| $t$ (s) | $\Delta t$ (s) |
|---|---|
| 0.015896440 | --- |
| 0.035876359 | 0.019979919 |
| 0.055877117 | 0.020000758 |
| 0.075877874 | 0.020000758 |
| 0.095878632 | 0.020000758 |
| 0.115879389 | 0.020000757 |
| 0.135880147 | 0.020000758 |
| 0.155880905 | 0.020000758 |
| 0.175881662 | 0.020000757 |

## Final remarks

In summary, we present a new simple experimental setup for demonstrating beat phenomenon. We have combined two amplitude-modulated light beams on a solar cell using two smartphones as signal generators and a third smartphone as an oscilloscope to visualize the resulting wave beats. A very good agreement is obtained between the modelling and the experimental result. It is an innovative approach to bring physics experimentation to the students and discover the potential possibilities of smartphones in basic physics courses.

## Acknowledgments

The authors would like to thank the Instituto de Ciencia de la Educación (Institute of Education Sciences) at the Universitat Politècnica de València (Technical University of Valencia), Spain, for its support to the teaching innovation group MSEL.

## References

1. P. Gluck, "You can illustrate beats without the usual tuning forks," *Phys. Educ.* 39 241, 2004
2. T. B. Greenslade, "Beats produced by a moving tuning fork," *Phys. Teach.* **31**, 443 (1993); https://doi.org/10.1119/1.2343837
3. A. Ganci and S. Ganci, "The simplest demonstration on acoustic beats," *Phys. Teach.* **53**, 32 (2015); https://doi.org/10.1119/1.4904239
4. J. Kuhn, P. Vogt and M. Hirth, "Analyzing the acoustic beat with mobile devices," *Phys. Teach*. **52**, 248 (2014); https://doi.org/10.1119/1.4868948
5. M. Osorio, C. J. Pereyra, D. L. Gau and A. Laguarda, "Measuring and characterizing beat phenomena with a smartphone," *Eur. J. Phys.* **39** (2018) 025708 (10pp), https://doi.org/10.1088/1361-6404/aa9034
6. M. H. Giménez, I. Salinas, J. A. Monsoriu, J. C. Castro-Palacio, "Direct visualization of mechanical beats by means of an oscillating smartphone," *Phys. Teach.* **55**, 424 (2017); https://doi.org/10.1119/1.5003745
7. C. A. Gaffney and D. Kagan, "Beats in an oscillator near resonance," *Phys. Teach*. **40**, 405 (2002); https://doi.org/10.1119/1.1517880
8. G. I. Opat, "The precession of a Foucault pendulum viewed as a beat phenomenon of a conical pendulum subject to a Coriolis force," *Am. J. Phys.* **59**, 822 (1991); doi: 10.1119/1.16729
9. H. J. Janssen and E. L. M. Flerackers, "Beats: an interference phenomenon," *Phys. Teach*. **17**, 188 (1979); https://doi.org/10.1119/1.2340177
10. M. McDonald, J. Ha, B. H. McGuyer, T. Zelevinsky, "Visible optical beats at the hertz level," *Am. J. Phys.* **82**, 1003 (2014); doi: 10.1119/1.4890502
11. H. Weltin, "Light beats," *Am. J. Phys.* **30**, 653 (1962); doi: 10.1119/1.1942153
12. R. Lane Reese, G. Y. Chang and D. L. Dupuy, "The oscillation of the synodic period of the Moon: A ''beating'' phenomenon," *Am. J. Phys.* **57**, 802 (1989); doi: 10.1119/1.15896
13. T. B. Greenslade, "Standing waves and beats displayed by foam packing," *Phys. Teach*. **26**, 396 (1988); https://doi.org/10.1119/1.2342555
14. I. Torriente-García, F. M. Muñoz-Pérez, A. C. Martí, M. Monteiro, J. C. Castro-Palacio, J. A. Monsoriu, "Experimenting with RC and RL series circuits using smartphones as function generators and oscilloscopes," *Rev. Bras. Ensino Fis.* **45**, e20230143 2023.
15. I. Torriente-García, A. C. Martí, M. Monteiro, C. Stari, J. C. Castro-Palacio, J. A. Monsoriu, "RLC series circuit made simple and portable with smartphones," *Phys. Educ.* **59**, 015016 2024.
16. Viera Sotware 2020 Physics toolbox, downloaded from Google Play on 20-2-2024.
17. S. Staacks, S. Hütz, H. Heinke and C. Stampfer, "Advanced tools for smartphone-based experiments: phyphox", *Phys. Educ.* **53**, 045009 2018.